\documentclass[a4paper]{article}
\usepackage{multicol}
\usepackage{epsfig}

\def\lesssim{\, \lower2truept\hbox{${<\atop\hbox{\raise4truept\hbox{$\sim$}}}$}\,}
\def\gtrsim{\,\lower2truept\hbox{${> \atop\hbox{\raise4truept\hbox{$\sim$}}}$}\,}

\begin{document}

%\title{\large MODELS OF GALAXY EVOLUTION
%FROM THE UV TO THE SUB-MILLIMETER}

%\author{
%{\normalsize \bf G.L.\ Granato}\\
%\\
%{\normalsize Osservatorio Astronomico di Padova, Italy}\\
%\\
%{\normalsize \bf  L.\ Silva,  L.\ Danese}\\
%\\
%{\normalsize SISSA, Trieste, Italy}\\
%\\
%{\normalsize \bf A.\ Bressan}\\
%\\
%{\normalsize Osservatorio Astronomico di Padova, Italy}\\
%\\
%{\normalsize \bf A.\ Franceschini, C.\ Chiosi}\\
%\\
%{\normalsize Dipartimento di Astronomia di Padova, Italy}\\
%}
%\date{}

%\maketitle

\centerline{\large MODELS OF GALAXY EVOLUTION
FROM THE UV TO THE SUB-MILLIMETER}
\bigskip
\centerline{\normalsize \bf G.L.\ Granato}
\medskip
\centerline{\normalsize Osservatorio Astronomico di Padova, Italy}
\medskip
\centerline{\normalsize \bf  L.\ Silva,  L.\ Danese}
\medskip
\centerline{\normalsize SISSA, Trieste, Italy}
\medskip
\centerline{\normalsize \bf A.\ Bressan}
\medskip
\centerline{\normalsize Osservatorio Astronomico di Padova, Italy}
\medskip
\centerline{\normalsize \bf A.\ Franceschini, C.\ Chiosi}
\medskip
\centerline{\normalsize Dipartimento di Astronomia di Padova, Italy}
\bigskip

\begin{multicols}{2}

\centerline{ABSTRACT}
\medskip
We present the first results of a detailed modeling of chemical and
photometric evolution of galaxies including the effects of a dusty interstellar
medium.

A chemical evolution code -- including infall of primeval gas, thermal and gravitational energy balance, and a galactic wind - follows the SF rate, the gas fraction and the metallicity, basic ingredients for the stellar population synthesis. The latter is performed with a grid of integrated spectra for single stellar populations (SSP) of different ages and metallicities, in which the effects of dusty envelopes around AGB stellar are included.

The residual fraction of gas in the galaxy is divided into two phases: the star forming molecular clouds and the diffuse medium. The relative amount is a model parameter. The molecular gas is sub-divided into clouds of given mass and radius: it is supposed that any SSP is borne within the cloud and progressively escapes it. The emitted spectrum 
of the star forming molecular clouds is computed with a radiative transfer code.  The diffuse dust emission (cirrus) is derived by describing the galaxy as an axially symmetric system, in which the local dust emissivity is consistently calculated as  a function of the  local field intensity due to the stellar component. Effects of very small grains, subject to temperature fluctuations, as well as PAH are included.

The model is compared and calibrated with available data of starburst galaxies in the
local universe, in particular new broad--band and spectroscopic ISO observations. It will be a fundamental tool to extract information on global evolution of the stellar component and of ISM of galaxies from observations covering four decades in $\lambda$. The combination of observations from HST, Keck, ISO and ground-based optical, IR and sub-mm telescopes already provided several objects at relevant $z$ with spectral information on this large $\lambda$ range. Their number will increase after the completion of ISO surveys, and will burst when SIRFT, FIRST, Planck Surveyor and NGST will operate.

\section{Introduction}

Several dusty environments must be taken into account in order to properly understand 
the IR and sub-mm properties of galaxies: (i) dust in interstellar HI clouds heated by the 
global radiation field of the galaxy (the `cirrus' component), (ii) dust associated with star forming molecular clouds and HII regions and (iii) circumstellar dust shells produced by the windy final stages of stellar evolution.

A satisfactory inclusion of these effects into codes following the chemical and spectrophotometric evolution of galaxies is still lacking, due both to the complexity as well as to the uncertainties of the physical situation, which can be hopefully clarified exploiting the full information contained in spectral energy dustributions (SEDs) ranging from the far-UV to the sub-millimeter. 

\section{Model description}
\subsection{Chemical model}
The chemical evolution is simulated through the code described by Bressan et al. (1994) 
and Tantalo et al.\ (1996) for elliptical galaxies, with a few modifications to
allow the description of other systems such as disks and starbursting objects.

The essential characteristics are summarized here, but we refer the reader to the above mentioned papers for more details. The code describes one-zone models with two components: luminous matter is embedded in a halo of dark matter whose presence affects only the gravitational potential of the galaxy and the binding energy of the gas.  It includes the infall of primordial gas, in order to simulate the collapse phase of galaxy formation and,  when required by the astrophysical situation under study,  galactic winds, whose occurrence is set by the balance between the thermal energy stored in the interstellar gas due to type I and II supernovae and the binding energy of the gas. In the original code, the adopted star formation rate (SFR) is a Schmidt-type law, i.e. proportional to some power (between 1 and 2) of the available gas mass: 
%\begin{equation
$
\Psi(t)=\nu \, M_{g}(t)^{k}
$
%\label{equ_SFR}
%\end{equation}
where $\nu$, the efficiency of SF, is the inverse star formation time scale when $k=1$, our usual choice. However, to simulate more general situations than spheroidal galaxies, we added considerable flexibility regarding the specification of the SFR, such as one or more bursts of star formation superimposed to the general smooth SFR. As for the IMF we used the usual Salpeter recipe: $\Phi(M)\propto M^{-x}$, with $x=2.35$, normalized by setting the fraction of total mass in the IMF above $1\,\rm{M}_{\odot}$, the minimum mass contributing to the enrichment of a galaxy ISM. 

The code has been successfully tested against nearby spheroidal galaxies (see the above references), for which it was originally designed, but it can simulate different types 
of galaxies with our modifications and/or an appropriate choice of the parameters (e.g. a quiet spiral-type evolution is well mimicked with a low $\nu$ and/or a long infall time scale). 

\typeout{(Bressan A., Chiosi C., \& Fagotto F., 1994, ApJS, 94, 63)}
\typeout{(Tantalo R., Chiosi C., Bressan A., \& Fagotto F., 1996, A\&A, 311, 361)}

\subsection{Photometric model} % stelle + gas

\subsubsection*{Synthesis of starlight spectrum}
The library of isochrones for the single stellar populations (SSPs),
the building-blocks of galaxy models, is based on the Padua stellar models, with a
major difference, consisting in the computation of  the effects of dusty envelopes 
around AGB stars (see below). The SSPs span a wide range in age, from 1 Myr to 20 Gyr, and in metallicity, $Z=0.08,0.2,0.5$, to realistically reproduce the mix of age and composition of the stellar content of galaxies. The spectral synthesis technique, for the starlight alone, consists in summing up the spectra of each stellar generation, provided by the SSP of the appropriate age and metallicity, weighted by the SFR at time of the stars birth. In this work the dust in the ISM is kept into account, which means that the spectrum of each SSP must be filtered through the dust, as described below, before performing the integration over the evolutionary  history of the galaxy.

\subsubsection*{Emission from circumstellar dust}
Effects of dust in the envelopes of Mira and OH stars are usually neglected in the synthesis of a composite population. While this can be justified in very old systems, it is not the case for intermediate age population clusters, whose brightest tracers are indeed the asymptotic giant branch stars. To overcame this limitation, we computed new isochrones in which, along the AGB, a suitable dusty envelope is assumed to surround the star. For this envelope the radiative transfer is solved by means of the code described by Granato \& Danese (1994). The envelope parameters, in  particular its optical thickness, which is straightforwardly linked to the mass-loss rate and to the expansion velocity, are derived as a function of basic stellar parameters (mass, luminosity and radius) either from hydrodynamic model results or from empirical relations. A detailed description of the adopted procedure can be found in Bressan, Granato \& Silva (1997).

\subsubsection*{Galaxy dust emission}

The dust content of the model galaxy is fixed by the residual gas, as provided by the star formation history, and by the chemical abundance of Carbon and Silicon. Starlight extinction from dust in the UV-optical spectral range and the resulting IR emission are computed under the assumption of a two-phase ISM, consisting of a molecular component, wherein new stars are born, and a diffuse one. The relative amount of molecular and diffuse gas is a fundamental model parameter.

{\it Molecular component.}
It is known that molecular clouds are the sites of star formation, and that stars begin their life embedded in dusty dark clouds. Thus the first evolutionary stages are hidden at optical wavelengths and a significant fraction, if not all, of the energy of young stellar objects is reprocessed by dust and radiated in the IR. The powerful stellar winds and outflows, and the ionizing flux from massive stars, all contribute to the destruction of the molecular clouds in a time scale comparable to the lifetime of OB stars,
$\sim 10^{6}-10^{7}$ yrs. Thus stars gradually get rid of their parent gas and become visible at optical wavelengths.

\typeout{(Lada 1993, in ``Molecular clouds and star formation'')}

This evolution is simulated as follows. The molecular gas is sub-divided into spherical clouds of assigned mass and radius, for which we adopt typical values found in GMCs, $\sim 10^{6}\; \rm{M}_{\odot}$ and $\sim 10-50$ pc respectively. \typeout{(Myers 1993, ``Molecular clouds and star formation'')} It is then supposed that each generation of stars, represented in our scheme by a SSP, is born within the cloud and progressively escapes it. This is mimicked by decreasing linearly with its age the fraction of SSP energy radiated inside the cloud. The proportionality coefficient of this relation is a model parameter, connected with the outflow velocity of young stars from their parent cloud and  fixing the fraction of blue light that can escape the starbursting region. The starlight which illuminates the cloud is approximated as a single central source. The cloud optical depth is fixed by the cloud mass and radius and by the dust-to-gas ratio. The emerging spectrum is obtained by solving the radiative transfer through the cloud with the code by Granato \& Danese (1994).

{\it Diffuse component - cirrus.}
Before escaping the galaxy, the light arising from stellar generations older than some $\sim$ Myr, no more embedded in the molecular clouds, as well as photons emitted by the young stars-molecular gas systems, further interacts with the dust present in the diffuse gas component. The twofold effect of this interaction, the dimming of the stellar light by extinction and the consequent dust emission, is computed by describing the galaxy as an axially symmetric system, subdivided into volume elements; for each of them the dust emission as a function of the local radiation field and the extinction by the dust column density are computed. The volume integral over the galaxy yelds the residual stellar spectrum and the diffuse dust emission.

The radial dependence of the stellar and gas densities are (independently) described with a King-type or an exponential profile. The code can treat also density distributions with a polar angle dependence, as adequate for flattened systems.

In the present version, it is assumed that the diffuse dust emission is never 
self-absorbed, so that a radiative transfer calculation is not performed. In the near future we plan to overcame this limitation which however is not severe for most interesting situations.

The dust consists of big grains in thermal equilibrium with the radiation field, small transiently heated grains and PAH molecules.  Once the bathing radiation field is specified, the detailed computation of the emissivity from each component is performed  following standard treatments (Guhathakurta \& Draine 1989, Puget et al.\ 1985).

\section{Results}

A few examples, showing the capability of our model to reproduce the data on starbursting galaxies in the local universe, are discussed below (see Fig.\ 1).

\subsection{M 82}
In the prototype starburst galaxy M82 the burst is probably triggered by the interaction with M81. Thanks to the relative  proximity of this system ($D = 3.25$ Mpc) a wealth of data do exist, allowing a well sampled full coverage of the SED at different angular resolutions. The fit is obtained by evolving for 10 Gyr a closed system with an initial barionic mass of $5 \, 10^{10} \, \rm{M}_\odot$. The assumed efficiency provides a SFR smoothly decreasing from 24 at $t=0$ to about 1 $\rm{M}_\odot/\rm{yr}$ at $t=10$ Gyr. To this gentle star formation history, which leaves a gas fraction of 0.034 and an almost solar mean metallicity, we have superposed a burst dissipating about half of the residual gas in the last $10^8$ yr. Thus we end with a total gas mass of $10^9 \, \rm{M}_\odot$, $4 \%$ of which is ascribed to the molecular component, organized in clouds with $M=10^{6} \, \rm{M}_\odot$ and $r=16$ pc. It is worth noticing that these clouds reprocess almost completely the starlight due to the burst. The system is assumed to follow a King profile with $r_c=200$ pc and the adopted value of the gas to dust mass ratio is 200.

\subsection{NGC 6090}
This strongly interacting galaxy at 175 Mpc ($\rm{H}_o=50$ km/s/Mpc) has been observed by ISO from 2.5 to 200 $\mu$m (Acosta-Pulido et al.\ 1996). The SED resulting from the combination of  this data with previously published optical photometry is nicely reproduced by our model. The differences in the parameters with respect to M82, apart from an up-scale in involved barionic mass ($9.4 \, 10^{11} \, \rm{M}_\odot$), are aimed at enhancing the diffuse dust emission at $\lambda \gtrsim 100 \, \mu$m, which, as noticed by Acosta-Pulido et al., is not reproduced by published starburst models. Thus the pre-burst SFR has been adjusted to leave a greater gas fraction (0.07), 20 \% of which is processed by the burst. The molecular star-forming clouds accounts only for 0.5 \% of the gas left by the burst. The dust masses in the cirrus and in the star-forming regions, once converted to the same H$_o$, are in excellent agreement with the estimates given by Acosta-Pulido et al..

\subsection{Arp 220}

Arp 220 is an archetypal ultraluminous infrared galaxy (ULIRG), most likely the result a recent merging between two gas-rich galaxies. In this object there is evidence of both starburst as well as Seyfert activity, but recent ISO spectroscopy leads to the conclusion that the IR luminosity is primarily powered by starburst (Sturm et al.\ 1996). The proposed fit for this objects differs from the previous ones for the fact that the burst is much stronger, processing as much as 17\% of the total barionic mass, which is $5 \, 10^{11} \, \rm{M}_\odot$, and for the high fraction 20\% of the residual gas in star-forming clouds. The mass in  molecular gas is therefore $\sim 2 \, 10^{10} \, \rm{M}_\odot$, in agreement with CO estimates (Scoville et al.\ 1991). As a result, the IR and sub-millimetric emission is everywhere dominated by dust associated with star forming regions.

\section{Conclusions}

The above examples show not only the model capability in matching  the spectral behavior of galaxies in a wavelength range as wide as 4 decades from the far-UV to the sub-mm, but also the accuracy in reproducing features such as those due to PAHs and silicates, which are relevant diagnostics of the ISM. 

Since chemical enrichment was a rapid process in normal elliptical and at least early type spiral galaxies, we expect that dust played a role soon after the onset of star formation. Thus only the multiband approach of these models is fully adequate to investigate spectra and evolution of galaxies, and well suited for already or soon available data from HST, Keck, ISO and ground-based optical, IR and sub-mm telescopes.
The model can be exploited to propose crucial tests for the theories of galaxy formation and evolution (e.g.\ counts in specific bands, observations of spectral features), to be performed with the next generation of instruments (SIRFT, FIRST, Planck Surveyor and NGST).

\bigskip

\centerline{REFERENCES}
\medskip

\noindent Acosta-Pulido, J.A., Klaas, U., Laureijs, R.J., et al., 1996, A\&A 315, L21

\noindent Bressan, A., Chiosi, C.\, Fagotto, F., 1994, ApJS, 94, 63

\noindent Bressan, A., Granato, G.L., Silva, L., 1997, in preparation

\noindent Granato, G.L.,  Danese, L., 1994, MNRAS, 268, 235

\noindent Guhathakurta, P., Draine, B.T., 1989, ApJ, 345,230

\noindent Puget, J.L., Leger, A., Boulanger, F., 1985, A\&A, 142, L19

\noindent Scoville, N.,Z., Sargent, A.I., Sanders, D.B., Soifer, B.T., 
1991, ApJ, 366, L5 

\noindent Sturm, E., Lutz, D., Sternberg, A., et al., 1996, A\&A, 315, L133

\noindent Tantalo, R., Chiosi, C., Bressan, A., Fagotto, F., 1996, A\&A, 311, 361

\end{multicols}

\begin{figure}
\vspace*{0pt}
\hspace*{50pt}
\epsfig{file=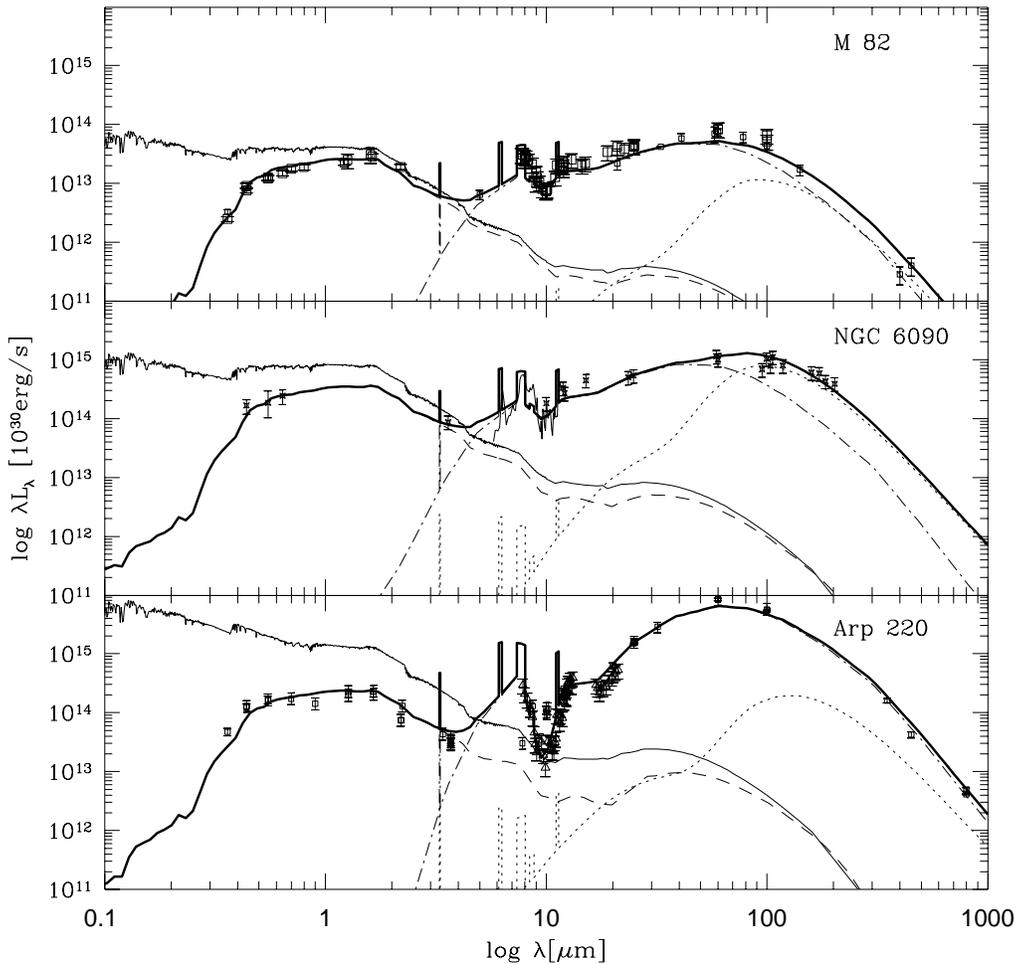,width=14.0cm}
\vspace*{0pt}

\caption{Fits to the SEDs of three starburst galaxies from the UV to the sub-mmm. Thin solid line: synthetic starlight spectrum without the effects of galactic dust; long dashed line: previous one with only extinction effects included; short dashed line: diffuse dust (cirrus) component; dot dashed line: molecular star-forming clouds; thick solid line: total;
}
\label{figtri}
\end{figure}

\end{document}